\documentclass[letter,desactivate]{aa} 

\usepackage{natbib}
\bibpunct{(}{)}{;}{a}{}{,}

\usepackage{graphicx}
\usepackage{revsymb}	
\usepackage{amsmath}
\usepackage{txfonts}
\usepackage{amssymb}
\usepackage{geometry} 
\usepackage[parfill]{parskip} 
\usepackage{slantsc}
\usepackage{array}
\usepackage{amsfonts}
	
\usepackage{epstopdf}	
\usepackage{epsfig}	

\usepackage{url}

\usepackage[colorlinks=true,linkcolor=black, urlcolor=black, citecolor=black]{hyperref}

\usepackage{color}

\usepackage{xfrac}

\usepackage{sidecap}
\usepackage[font=small, labelfont=bf]{caption}
\usepackage{subcaption}

\usepackage{multirow}
\usepackage{booktabs}
\usepackage{threeparttable}

\begin{document}
   \title{Frequency separation ratios do not suppress magnetic activity effects in solar-like stars}
\author{J. B\'{e}trisey\inst{1} \and A. -M. Broomhall\inst{2} \and S.~N. Breton\inst{3} \and R.~A. Garc{\'\i}a\inst{4} \and H. Davenport\inst{2,5}\and O. Kochukhov\inst{1}}
\institute{Department of Physics and Astronomy, Uppsala University, Box 516, SE-751 20 Uppsala, Sweden\\email: 	\texttt{jerome.betrisey@physics.uu.se}
\and Centre for Fusion, Space and Astrophysics, Department of Physics, University of Warwick, Coventry CV4 7AL, UK
\and  INAF – Osservatorio Astrofisico di Catania, Via S. Sofia 78, 95123 Catania, Italy
\and Université Paris-Saclay, Université Paris Cité, CEA, CNRS, AIM, 91191 Gif-sur-Yvette, France
\and Blackett Laboratory, Imperial College London, London SW7 2AZ, United Kingdom}
\date{\today}

\abstract
{Asteroseismic modelling will play a critical role in future space-based missions such as PLATO, CubeSpec, and \textit{Roman}. Magnetic activity effects were typically neglected in asteroseismic modelling of solar-type stars, presuming that these effects could be accounted for in the parametrisation of the so-called `surface effects'. In the recent years, it was however demonstrated that magnetic activity can have a significant impact on the asteroseismic characterisation using  both forward and inverse techniques.}
{We investigated whether frequency separation ratios, which are commonly used to efficiently suppress surface effects, are also able to suppress magnetic activity effects.}
{Based on GOLF and BiSON observations of the Sun-as-a-star that were segmented into yearly overlapping snapshots, each offset by a quarter of a year, we performed asteroseismic characterisations using frequency separation ratios as constraints to measure the apparent temporal evolution of the stellar parameters and their correlation with the 10.7 cm radio flux, a solar activity proxy.}
{Frequency separation ratios do not suppress the effects of magnetic activity. Both $r_{01}$ and $r_{02}$ ratios exhibit a clear signature of the magnetic activity cycle. Consequently, when these ratios are employed as constraints in asteroseismic modelling, magnetic activity effects are propagated  to the stellar characterisation. Additionally, most stellar parameters correlate with the activity cycle, unlike the direct fitting of individual frequencies. These findings are consistent across both the GOLF and BiSON datasets.}
{Magnetic activity effects significantly impact asteroseismic characterisation within the current modelling framework, regardless of whether forward modelling or inverse methods are used. Moreover, standard techniques to suppress surface effects have proven ineffective against magnetic activity influences. If the latter are used, systematic uncertainties of 4.7\%, 2.9\%, and 1.0\% should be considered for the stellar age, mass, and radius, respectively. In preparation for future space-based photometry missions, it is therefore essential to enhance our theoretical understanding of these effects and develop a modelling procedure capable of accounting for or efficiently suppressing them.}


\maketitle

\defcitealias{Betrisey2023_AMS_surf}{JB23}
\defcitealias{Betrisey2024_MA_Sun}{JB24}
\defcitealias{Betrisey2025}{JB25}

\section{Introduction}
The convective motions in the outer layers of solar-type stars excite an extended set of stellar oscillations. By examining these oscillations, asteroseismology offers an unparalleled glimpse into the internal structure of stars, allowing us to determine their fundamental parameters with a high level of precision and accuracy. Accurate stellar models are crucial for understanding planetary system evolution and tracing the history of our own galaxy \citep[see e.g.][for reviews]{Chaplin&Miglio2013,Garcia&Ballot2019,Aerts2021}. Building on the success of missions such as CoRoT \citep[Convection, Rotation and planetary Transits;][]{Baglin2009}, \textit{Kepler} \citep{Borucki2010}, K2 \citep{Howell2014}, and TESS \citep[Transiting Exoplanet Survey Satellite;][]{Ricker2015}, asteroseismic modelling is set to play a pivotal role in upcoming space-based missions such as PLATO \citep[PLAnetary Transits and Oscillations of stars;][]{Rauer2024}, CubeSpec \citep{Bowman2022}, and Roman \citep{Huber2023}. Despite impressive progresses in asteroseismology and modelling of stellar evolution, discrepancies remain between observed data and theoretical models, leading to biases in stellar characterisation. This issue is increasingly becoming pressing given the stringent precision requirements of the PLATO mission (15\% in mass, 1-2\% in radius, and 10\% in age for Sun-like stars).

Three main challenges are impacting asteroseismic modelling: surface effects \citep[e.g.][]{Ball&Gizon2017,Nsamba2018,Jorgensen2020,Jorgensen2021,Cunha2021,Betrisey2023_AMS_surf}, stellar magnetic activity \citep[e.g.][]{Garcia2010,Broomhall2011,Santos2018,PerezHernandez2019,Santos2019_sig,Santos2019_rot,Howe2020,
Thomas2021,Santos2021,Santos2023,Betrisey2024_MA_Sun,Betrisey2025}, and the physical ingredients used in stellar models \citep[e.g.][]{Buldgen2019f,Farnir2020,Betrisey2022}. We refer for example to \citet{Betrisey2024_phd} for an in-depth discussion of these challenges. Similar to surface effects, magnetic activity effects induce shifts in the observed oscillations frequencies \citep[see e.g.][for a review]{Broomhall&Nakariakov2015}. The origin of these shifts remains unclear but two main hypotheses emerge from the dedicated literature: magnetic field variations in the near-surface layers \citep[e.g.][]{Howe2002, Baldner2009,Garcia2024} and structural changes in the acoustic cavity \citep[e.g.][]{Woodard&Noyes1985,Fossat1987,Libbrecht&Woodard1990,Kuhn1998,
Dziembowski&Goode2005,Basu2012}. Magnetic activity effects are propagated to the asteroseismic characterisation, regardless of whether forward modelling \citep{Creevey2011,PerezHernandez2019,Thomas2021,Betrisey2024_MA_Sun} or inverse methods are used \citep{Betrisey2025}.

Given the non-negligible impact of magnetic activity on asteroseismic characterisation within the current modelling framework, this letter investigates whether conventional methods for effectively suppressing surface effects, which also induce frequency shifts, can mitigate the influence of magnetic activity. In Sect.~\ref{sec:influence_MA_observed_ratios}, we assess the impact of magnetic activity on frequency separation ratios, which are the observational constraints employed in these methods. In Sect.~\ref{sec:influence_MA_characterisation}, we examine whether the effects of magnetic activity are transmitted to the asteroseismic characterisation. Finally, we present the conclusions of our study in Sect.~\ref{sec:conclusions}.

\section{Influence of magnetic activity on observed frequency separation ratios}
\label{sec:influence_MA_observed_ratios}
Similar to \citet{Betrisey2024_MA_Sun} and \citet{Betrisey2025} (hereafter JB24 and JB25), we based our analysis on two independent datasets of Doppler velocity observations of the Sun-as-a-star: BiSON \citep[Birmingham Solar Oscillations Network;][]{Davies2014,Hale2016} and GOLF \citep[Global Oscillations at Low Frequencies;][]{Gabriel1995}. We recall that these datasets correspond to solar observations that were segmented into yearly overlapping snapshots, each offset by a quarter of a year. The original datasets cover solar cycles 23 and 24. In this study, we revised the BiSON dataset using a more sophisticated peak-bagging algorithm (see Appendix~\ref{sec:app:observational_data}) and extending it to include the first half of the maximum of cycle 25. For this section, similarly to \citet{Salabert2015}, we also divided the frequency range into three sub-ranges (see Table~\ref{tab:frequency_ranges}) to measure whether magnetic activity effects are affecting more certain regimes of the frequency separation ratios $r_{01}$ and $r_{02}$, defined in Appendix~\ref{sec:app:supplementary_data_observed_ratios}.

\begin{figure}
\centering
\begin{subfigure}[b]{.4\textwidth}
  \includegraphics[width=.99\linewidth]{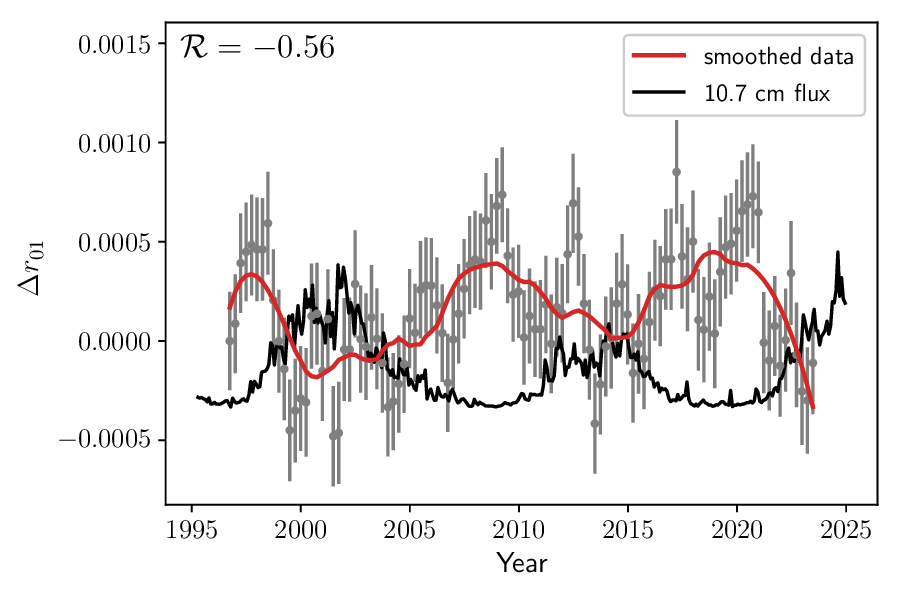} 
\end{subfigure}
\begin{subfigure}[b]{.4\textwidth}
  \includegraphics[width=.99\linewidth]{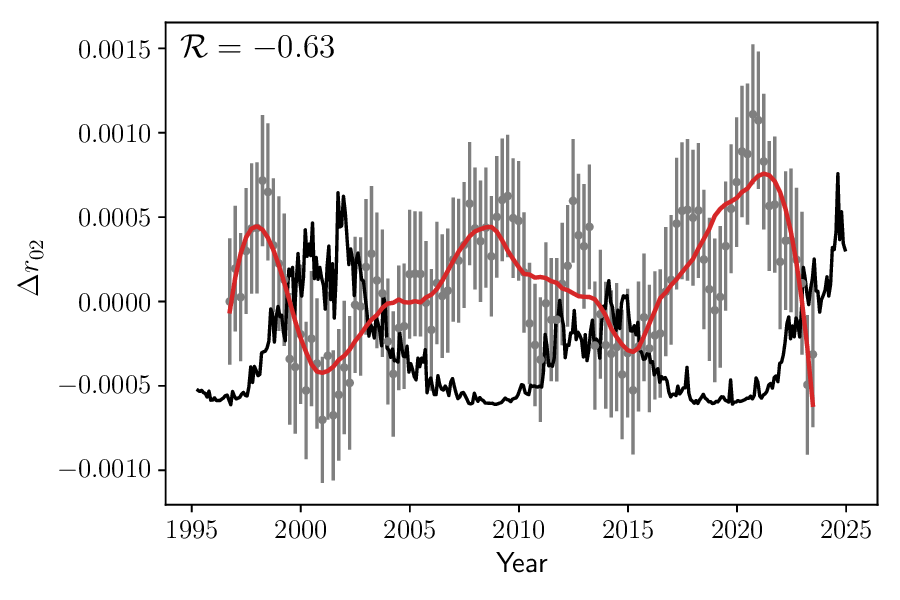} 
\end{subfigure}
\caption{Influence of the magnetic activity cycle on frequency separation ratios $r_{01}$ and $r_{02}$ observed by BiSON. The grey data represent the averaged shift due to magnetic activity. The full frequency range was considered ($n=14-24$). The red line is obtained by smoothing the grey data with a Savitzky-Golay filter. The 10.7 cm radio emission flux, a proxy of the solar cycle, is denoted in black. The correlation with the activity proxy is measured with the Spearman rank coefficient $\mathcal{R}$.}
\label{fig:imprint_observed_ratios_main_text}
\end{figure}

In Fig.~\ref{fig:imprint_observed_ratios_main_text}, we present the temporal evolution of the frequency separation ratios $r_{01}$ and $r_{02}$. To emphasize the impact of the magnetic activity cycle, we employed a method commonly used in the literature for individual frequencies \citep[see e.g.][]{Broomhall&Nakariakov2015}, which involves calculating the shift relative to a reference value, in our case, the value from the first snapshot. We then computed an averaged shift by averaging the shifts across several radial orders $n$. For instance, the averaged $r_{01}$ shift at time $t_j$ is determined as follows:
\begin{align}
\Delta r_{01}(t_j) = \frac{1}{n_{max}-n_{min}+1}\sum_{n=n_{min}}^{n_{max}}\left(r_{01}(t_j,n)-r_{01}(t_0,n)\right),
\end{align}
where $j$ is the snapshot number (e.g., 0 to 93 for GOLF datasets). The radial-order boundaries $n_{min}$ and $n_{max}$ are specified in Table~\ref{tab:frequency_ranges}. As shown in Fig.~\ref{fig:imprint_observed_ratios_main_text}, both ratio shifts exhibit a significant (anti-)correlation with the solar activity cycle proxy, the 10.7 cm radio emission flux\footnote{\url{https://www.spaceweather.gc.ca/}}, clearly indicating an influence of the activity cycle on the observed frequency separation ratios. Additionally, we observe that this imprint becomes more pronounced with increasing radial order (see Appendix~\ref{sec:app:supplementary_data_observed_ratios}), as expected from the literature \citep{Chaplin2005,Thomas2019,Thomas2021}.

\begin{table}
\centering
\caption{Frequency ranges considered in the computation of the averaged frequency separation ratio shift due to magnetic activity.}
\resizebox{0.99\linewidth}{!}{
\begin{tabular}{lcccc}
\hline \hline
\multirow{2}{*}{Frequency range} & \multicolumn{2}{c}{BiSON} & \multicolumn{2}{c}{GOLF} \\
 \cmidrule[0.4pt](lr){2-3} \cmidrule[0.4pt](l){4-5}
& $n$ & $\nu$ ($\mu$Hz) & $n$ & $\nu$ ($\mu$Hz) \\
 \hline
Full & $14-24$ & $2020-3510$ & $13-26$ & $1890-3780$ \\
Low-frequency& $14-17$ & $2020-2560$ & $13-17$ & $1890-2560$ \\
Medium-frequency & $18-21$ & $2560-3100$ & $18-21$ & $2560-3100$ \\
High-frequency & $22-24$ & $3100-3560$ & $22-26$ & $3100-3780$ \\
\hline
\end{tabular}}
\label{tab:frequency_ranges}
\end{table}

\section{Influence of magnetic activity on asteroseismic characterisation}
Similar to the effects of magnetic activity, surface effects, which are primarily linked to the inaccurate treatment of convection by the mixing-length theory in the upper stellar layers where acoustic oscillations are excited, also induce frequency shifts. Over the past two decades, significant efforts have been dedicated to understanding these surface effects, leading to the development of several approaches to mitigate the frequency shifts they cause. Among these, two main strategies are commonly employed: semi-empirical corrections to model the shifts \citep[e.g.][]{Kjeldsen2008,Ball&Gizon2014,Sonoi2015} and methods that efficiently damp these shifts \citep[e.g.][]{Roxburgh&Vorontsov2003,Oti2005,Roxburgh2016}. As discussed in the introduction, the former approach does not prevent magnetic activity effects from biasing the asteroseismic characterisation. In this section, we investigate whether the second approach yields better results.

One of the primary mitigation strategies involves using frequency separation ratios instead of individual frequencies, as these ratios have been shown to effectively suppress surface effects \citep{Roxburgh&Vorontsov2003,Oti2005}. To measure the temporal evolution of the asteroseismic characterisation based on this approach, we performed asteroseismic characterisations of each snapshot using the $r_{01}$ and $r_{02}$ ratios, the mean density determined by \citetalias{Betrisey2025} (hereafter inverted mean density), and the same spectroscopic constraints as \citetalias{Betrisey2024_MA_Sun}. We utilised a grid-based global minimisation algorithm that combines an interpolation scheme with a high-resolution grid of stellar models to extensively explore the parameter space and infer robust stellar parameters. For a detailed description of the minimisation process, we refer to \citet{Betrisey2023_AMS_surf}. Following \citetalias{Betrisey2025}, the correlation with the activity proxy is then investigated with the Spearman rank correlation coefficient $\mathcal{R}$ \citep{Spearman1904}.

\begin{figure}
\centering
\begin{subfigure}[b]{.43\textwidth}
  \includegraphics[width=.99\linewidth]{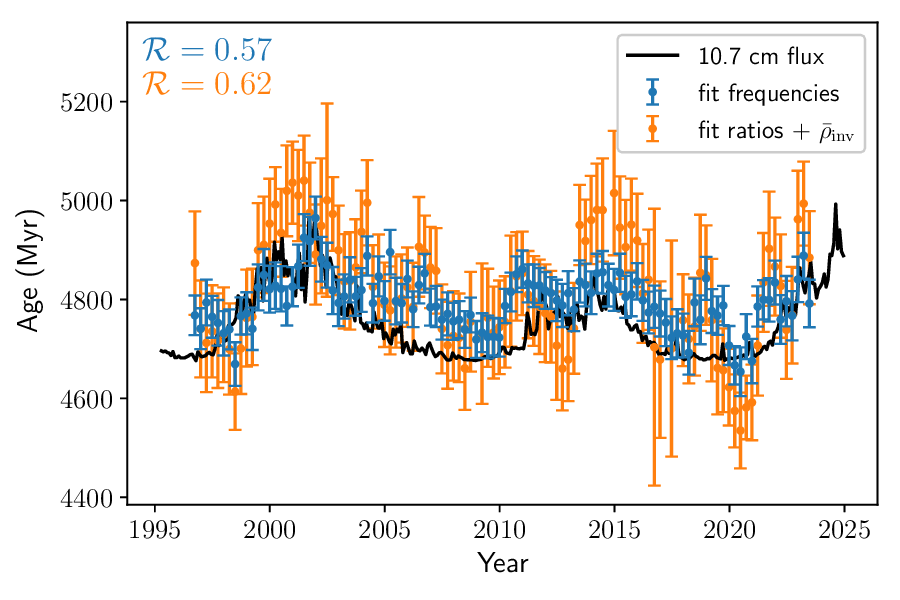} 
\end{subfigure}
\begin{subfigure}[b]{.43\textwidth}
  \includegraphics[width=.99\linewidth]{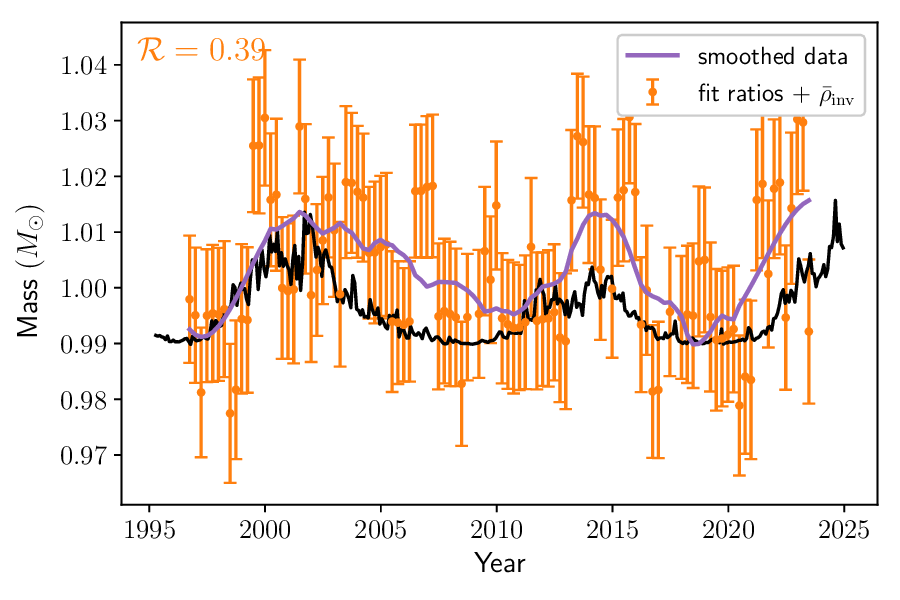} 
\end{subfigure}
\caption{Influence of the magnetic activity cycle on the asteroseismic age and mass based on BiSON data. The individual frequencies were used as constraints for the blue data, while the frequency separation ratios ($r_{01}$ and $r_{02}$) and the inverted mean density $\bar{\rho}_{\mathrm{inv}}$ were used for the orange data. The 10.7 cm radio emission flux, which serves as a proxy of the solar cycle, is denoted in black. The correlation with the activity proxy is measured with the Spearman rank coefficient $\mathcal{R}$. The purple line is obtained by smoothing the orange data with a Savitzky-Golay filter.}
\label{fig:impact_characterisation}
\end{figure}

In the upper panel of Fig.~\ref{fig:impact_characterisation}, we compare our results (in orange) for the asteroseismic age with those of \citetalias{Betrisey2024_MA_Sun}, who employed semi-empirical corrections (in blue). Stellar age is the most sensitive parameter to magnetic activity effects when semi-empirical corrections are used \citepalias{Betrisey2024_MA_Sun}. As demonstrated in the previous section, the observed ratios are sensitive to magnetic activity effects, and this signal is carried through to the asteroseismic characterisation. However, we observe a larger impact of magnetic activity effects, approximately 4.7\% compared to about 3\% with semi-empirical corrections. Furthermore, a constant bias of about 300 Myr is observed for the asteroseismic age. This bias is primary caused by the abundances used in the models of the high-resolution grid \citep[e.g.][]{Bonanno&Frohlich2015}, which are based on the determinations of \citet{Asplund2009}. This issue, which is more broadly known as the solar modelling problem \citep[see e.g.][]{Basu&Antia2008,Buldgen2019e}, has no influence on the results of our study.

Additionally, most stellar parameters show a weak to moderate imprint of the magnetic cycle (see Appendix~\ref{sec:app:supplementary_data_asteroseismic_characterisation}) with correlation coefficients $\mathcal{R}$ between 0.3 and 0.5, unlike results with semi-empirical corrections, where magnetic activity effects are limited to certain stellar variables such as age, mean density, and acoustic radius. Notably, stellar mass, displayed in the lower panel of Fig.~\ref{fig:impact_characterisation}, and radius show a significant correlation with the activity proxy (with $\mathcal{R}\simeq 0.4$). In this case, systematic uncertainties of about 2.9\% and 1.0\% should be associated with magnetic activity effects, respectively. We note that mass, age, and initial helium mass fraction $Y_0$ are free variables of the minimisation and are interconnected. The observed mass variation is thus not solely attributable to changes in age, but also to changes in $Y_0$. The modulation of $Y_0$ across different phases of the activity cycle is substantial and dominates the resulting mass trend.

As shown by \citetalias{Betrisey2024_MA_Sun} and implicitly assuming monotonically increasing magnetic effects, despite their inefficiency, semi-empirical corrections can partially absorb the magnetic activity signal in the free variables introduced to model surface effects, explaining this difference. With frequency separation ratios, surface effects are very efficiently suppressed by construction of the ratios, and no additional free variables are required in the minimisation process to account for these effects. Hence, the full activity signal is propagated to the asteroseismic characterisation. Although this may seem problematic at first, it presents a promising avenue. In the ratios space, surface effects and magnetic activity effects are disentangled, making it possible to define a semi-empirical correction to model magnetic activity effects in this constraint space, similar to how surface effects were modelled in the frequency space \citep{Kjeldsen2008,Ball&Gizon2014,Sonoi2015}.

The inverted mean density used as a constraint is also weakly sensitive to magnetic activity. To test which type of constraint transmits the most activity signal, we repeated the computations using a fixed inverted mean density, specifically the optimal value of \citetalias{Betrisey2025} based on the averaging of data over two full cycles, and found negligible differences. This indicates that the activity signal comes almost solely from the observed ratios. This is an interesting observation since the mean density determined through seismic inversions is typically used to improve stellar modelling, meaning no additional bias is introduced by this constraint concerning magnetic activity effects.

\label{sec:influence_MA_characterisation}

\section{Conclusions}
\label{sec:conclusions}
In this study, we explored whether standard techniques used to mitigate surface effects, specifically fitting frequency separation ratios using a forward modelling algorithm, could also suppress the effects of magnetic activity. We based our analysis on two independent datasets from GOLF and BiSON radial velocity observations of the Sun-as-a-star. In Sect.~\ref{sec:influence_MA_observed_ratios}, we assessed the impact of magnetic activity on the observed $r_{01}$ and $r_{02}$ ratios, and in Sect.~\ref{sec:influence_MA_characterisation}, we examined whether magnetic activity effects are propagated to the asteroseismic characterisation.

Our findings indicate that frequency separation ratios do not suppress the effects of magnetic activity. Both $r_{01}$ and $r_{02}$ ratios exhibit a clear signature of the magnetic activity cycle. Similar to observations of oscillation frequencies \citep[e.g.][and references therein]{Broomhall&Nakariakov2015}, the higher the radial order, the more the ratios are affected by magnetic activity. Consequently, when these ratios are used as constraints in asteroseismic modelling, magnetic activity effects propagate to the stellar characterisation. Additionally, most stellar parameters, including mass, age, radius, and initial helium mass fraction, correlate with the solar activity cycle. These findings are consistent across both the GOLF and BiSON datasets. This behaviour differs from when individual frequencies are used as constraints, where magnetic activity effects are partially absorbed by the surface effect prescription, significantly impacting only a limited number of stellar parameters, such as age, mean density, and acoustic radius \citep{Betrisey2024_MA_Sun,Betrisey2025}. Unfortunately, these parameters are among the primary stellar parameters of relevance for the PLATO mission \citep[e.g.][]{Rauer2024}. Moreover, the asteroseismic age is more sensitive to magnetic activity effects when ratios are used as constraints. In terms of systematic uncertainty, we observe an increase from about 3\% when individual frequencies are used as constraints \citep{Betrisey2024_MA_Sun} to about 4.7\% with frequency separation ratios. For the stellar mass and radius, systematic uncertainties of about 2.9\% and 1.0\% should be considered, respectively. These values are not negligible considering the PLATO precision requirements for Sun-like stars, especially for the stellar radius.

Within the current modelling framework, magnetic activity effects significantly impact asteroseismic characterisation, regardless of whether forward modelling or inverse methods are used \citep{Creevey2011,PerezHernandez2019,Thomas2021,
Betrisey2024_MA_Sun,Betrisey2025}. As demonstrated in this study, standard techniques to suppress surface effects have proven ineffective against magnetic activity influences. Since phase matching fitting is a variant of our methodology \citep{Roxburgh2016}, we expect similar results but it would be worthwhile to carry out a dedicated study to confirm this extrapolation. In preparation for future space-based photometry missions, it becomes essential to enhance our theoretical understanding of these effects and develop a modelling procedure capable of accounting for or efficiently suppressing them. In this context, we believe that frequency separation ratios may play a crucial role. Since these ratios are designed to efficiently damp surface effects \citep{Roxburgh&Vorontsov2003,Oti2005}, they allow for the disentanglement of surface and magnetic effects. It is therefore conceivable to develop semi-empirical corrections to damp magnetic activity effects in this constraint space, similar to what has been done for oscillation frequencies \citep{Kjeldsen2008,Ball&Gizon2014,Sonoi2015}.


\begin{acknowledgements}
We thank Dr. A.M. Amarsi for the fruitful discussions associated with the accessibility of the letter to non-specialised readers. J.B. acknowledges funding from the SNF Postdoc.Mobility grant no. P500PT{\_}222217 (Impact of magnetic activity on the characterization of FGKM main-sequence host-stars). A.-M.B. has received support from STFC consolidated grant ST/X000915/1. H.D. and A.-M.B. acknowledge support from the Royal Astronomical Society for the Undergraduate Summer Research Bursary. S.N.B acknowledges support from PLATO ASI-INAF agreement no. 2022-28-HH.0 "PLATO Fase D". R.A.G. acknowledges support from PLATO and GOLF CNES grants. O.K. acknowledges support by the Swedish Research Council (grant no. 2023-03667) and the Swedish National Space Agency.
\end{acknowledgements}


\bibliography{bibliography.bib}

\appendix

\section{Observational data}
\label{sec:app:observational_data}
We reused the GOLF datasets ($n\geq 12$ and $n\geq 16$) from \citet{Betrisey2024_MA_Sun,Betrisey2025}. We refer to these articles and to \citet{Garcia2005} and \citet{Breton2022_helio}, for a detailed explanation of the extraction of the oscillations frequencies from the GOLF observational timeseries. The method for fitting the BiSON data was based on that described in \citet{Broomhall2009_def}, with the key difference that the Affine Invariant Markov Chain Monte Carlo (MCMC) Ensemble sampler \texttt{emcee} \citep{Foreman-Mackey2013} was used instead of a hill-climbing algorithm. The process was very similar to that described in \citet{Santos2018}. To summarise, the modes were fitted in their respective $l=0/2$, $l=1/3$ pairs. An asymmetric Lorentzian was fitted to each visible peak as well as a flat background term, which is valid over the localized range close to the modes. For stability, the heights of the inner components ($m=0$ for $l=2$ and $|m|=1$ for $l=3$) were fixed using average values obtained by \citet{Broomhall2009_def}, which were obtained when fitting a much longer timeseries than those used in this study. While the widths of the peaks were a free parameter in the code, also for stability, the widths were set to be the same for each component observed in the respective $l=0/2$, $l=1/3$ pairs. Priors for the fits were based on the values obtained by \citet{Broomhall2009_def}. Outputs of the MCMC fitting routine were compared to those obtained by the routine used by \citet{Broomhall2009_def} and found to be consistent.

\section{Supplementary data for the observed frequency separation ratios}
\label{sec:app:supplementary_data_observed_ratios}
In Table~\ref{tab:characteristics_observed_shifts}, we present the characteristics of the average shifts induced by magnetic activity for the $r_{01}$ and $r_{02}$ ratios, as well as the frequencies of the radial order modes $\nu_0$. The frequency separation ratios were computed according to the definitions of \citet{Roxburgh&Vorontsov2003}. Specifically,
\begin{align}
r_{01}(n) &= \frac{\frac{1}{8}\left(\nu_{n-1,0}-4\nu_{n-1,1}+6\nu_{n,0}-4\nu_{n,1}+\nu_{n+1,0}\right)}{\nu_{n,1}-\nu_{n-1,1}}
\end{align}
and
\begin{align}
r_{02}(n) &= \frac{\nu_{n,0}-\nu_{n-1,2}}{\nu_{n,1}-\nu_{n-1,1}}.
\end{align}
The correlation between these average shifts and the activity proxy, the 10.7 cm radio emission flux, is assessed using the Spearman rank coefficient $\mathcal{R}$. We consider the data to be uncorrelated, or more precisely, negligibly correlated, when $|\mathcal{R}| < 0.25$. We introduce $\mathcal{A}$ as a proxy for the amplitude of the average shift, calculated by determining the difference between the maximum and minimum values of the data smoothed with a Savitzky-Golay filter \citep{Savitzky&Golay1964}. The primary purpose of this proxy is to facilitate comparisons between different frequency ranges of the same observable (i.e., $\Delta r_{01}$, $\Delta r_{02}$, and $\Delta\nu_0$).

\begin{figure*}
\resizebox{0.99\linewidth}{!}{
\begin{subfigure}[b]{.49\textwidth}
  \includegraphics[width=.99\linewidth]{Figures/bison_dr01_obs_all.eps} 
  \caption{\centering Full frequency range: 2020 - 3510 $\mu$Hz}
\end{subfigure}
\begin{subfigure}[b]{.49\textwidth}
  \includegraphics[width=.99\linewidth]{Figures/bison_dr02_obs_all.eps} 
  \caption{\centering Full frequency range: 2020 - 3510 $\mu$Hz}
\end{subfigure}
\begin{subfigure}[b]{.49\textwidth}
  \includegraphics[width=.99\linewidth]{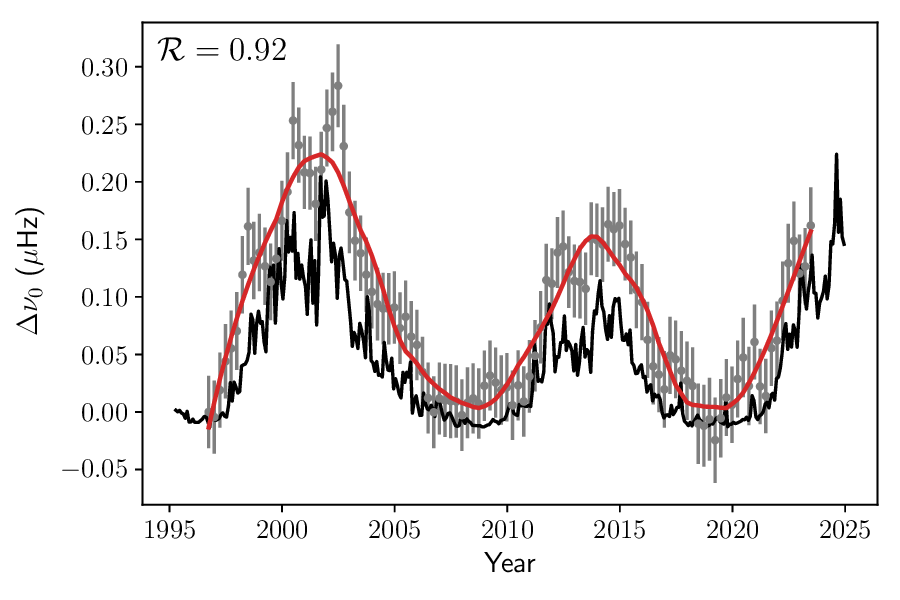} 
  \caption{\centering Full frequency range: 2020 - 3510 $\mu$Hz}
\end{subfigure}}
\resizebox{0.99\linewidth}{!}{
\begin{subfigure}[b]{.49\textwidth}
  \includegraphics[width=.99\linewidth]{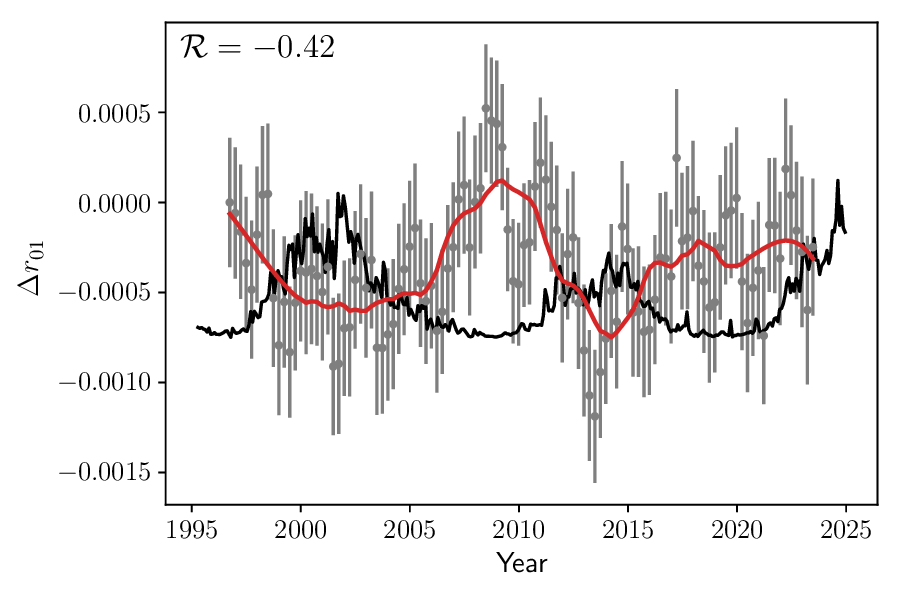} 
  \caption{\centering Low-frequency range: 2020 - 2560 $\mu$Hz}
\end{subfigure}
\begin{subfigure}[b]{.49\textwidth}
  \includegraphics[width=.99\linewidth]{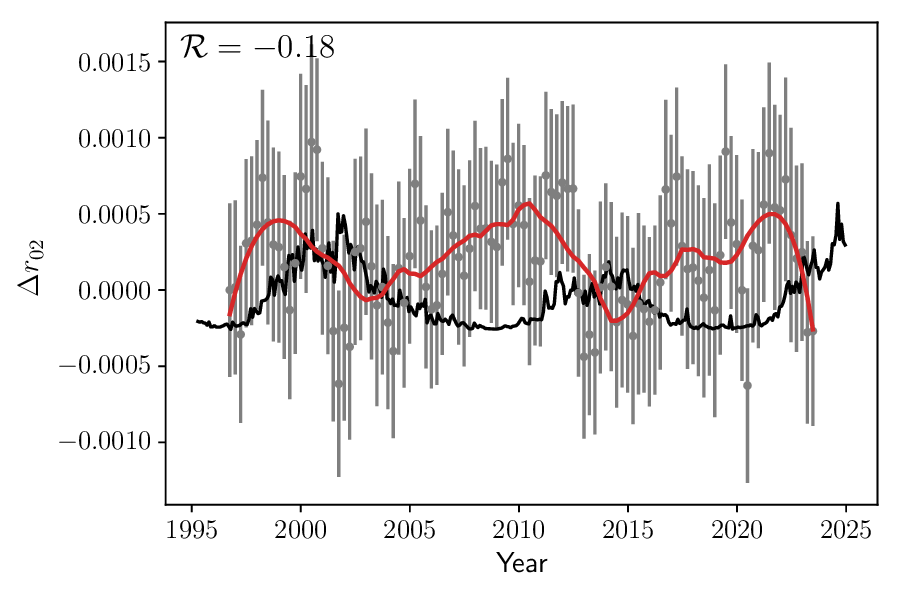} 
  \caption{\centering Low-frequency range: 2020 - 2560 $\mu$Hz}
\end{subfigure}
\begin{subfigure}[b]{.49\textwidth}
  \includegraphics[width=.99\linewidth]{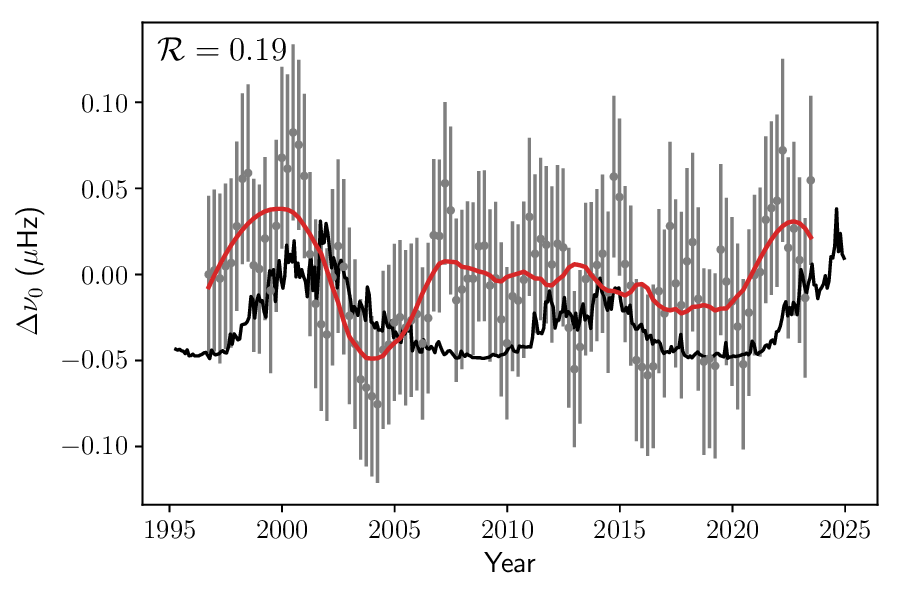} 
  \caption{\centering Low-frequency range: 2020 - 2560 $\mu$Hz}
\end{subfigure}}
\resizebox{0.99\linewidth}{!}{
\begin{subfigure}[b]{.49\textwidth}
  \includegraphics[width=.99\linewidth]{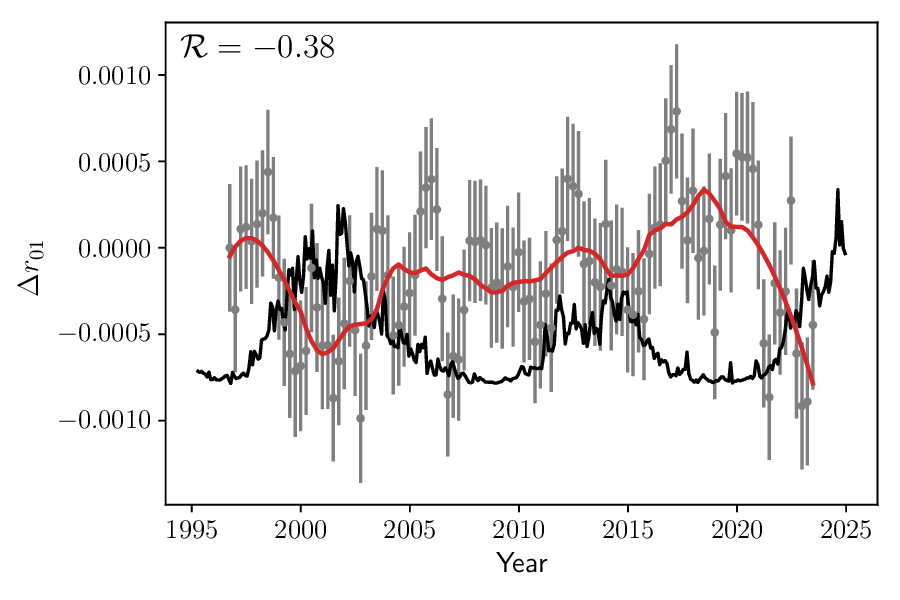} 
  \caption{\centering Medium-frequency range: 2560 - 3100 $\mu$Hz}
\end{subfigure}
\begin{subfigure}[b]{.49\textwidth}
  \includegraphics[width=.99\linewidth]{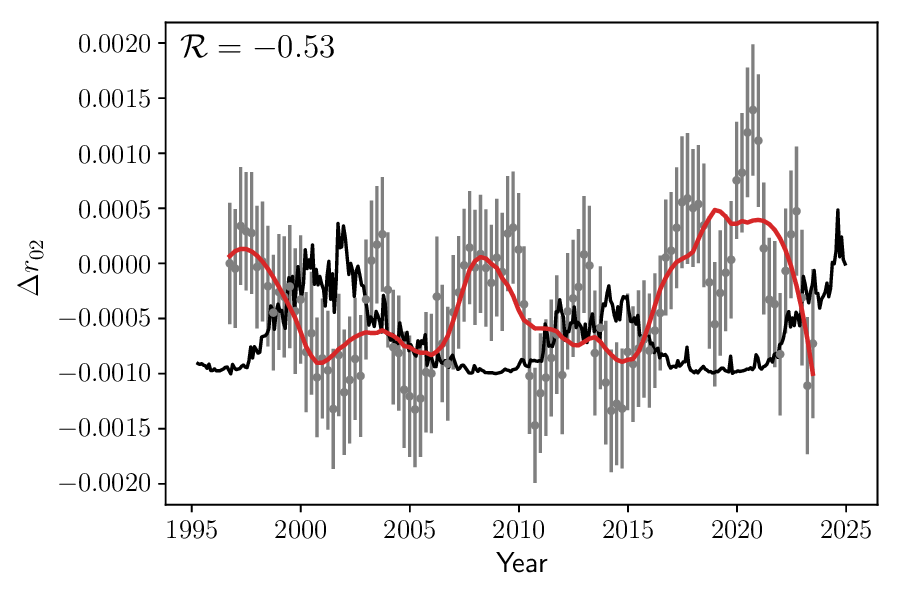} 
  \caption{\centering Medium-frequency range: 2560 - 3100 $\mu$Hz}
\end{subfigure}
\begin{subfigure}[b]{.49\textwidth}
  \includegraphics[width=.99\linewidth]{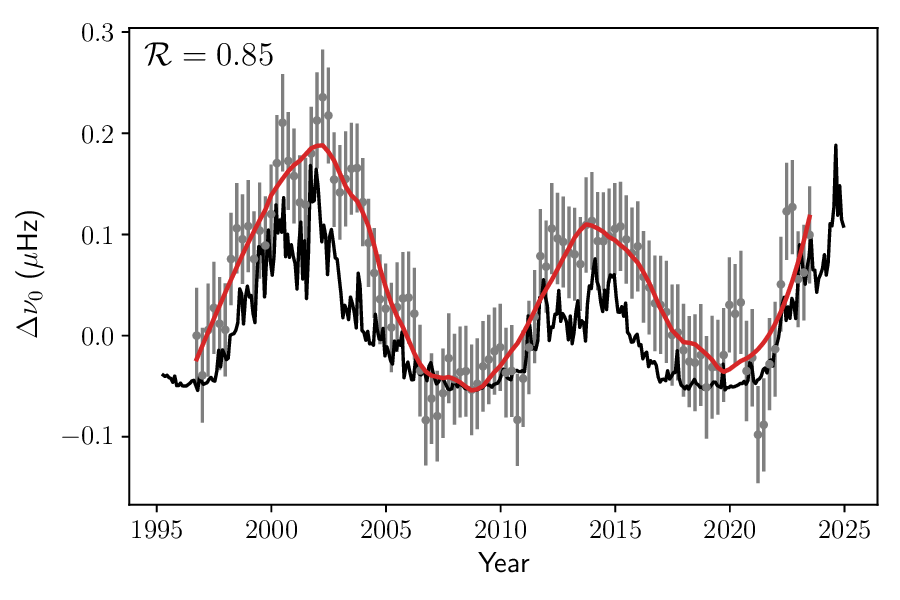} 
  \caption{\centering Medium-frequency range: 2560 - 3100 $\mu$Hz}
\end{subfigure}}
\resizebox{0.99\linewidth}{!}{
\begin{subfigure}[b]{.49\textwidth}
  \includegraphics[width=.99\linewidth]{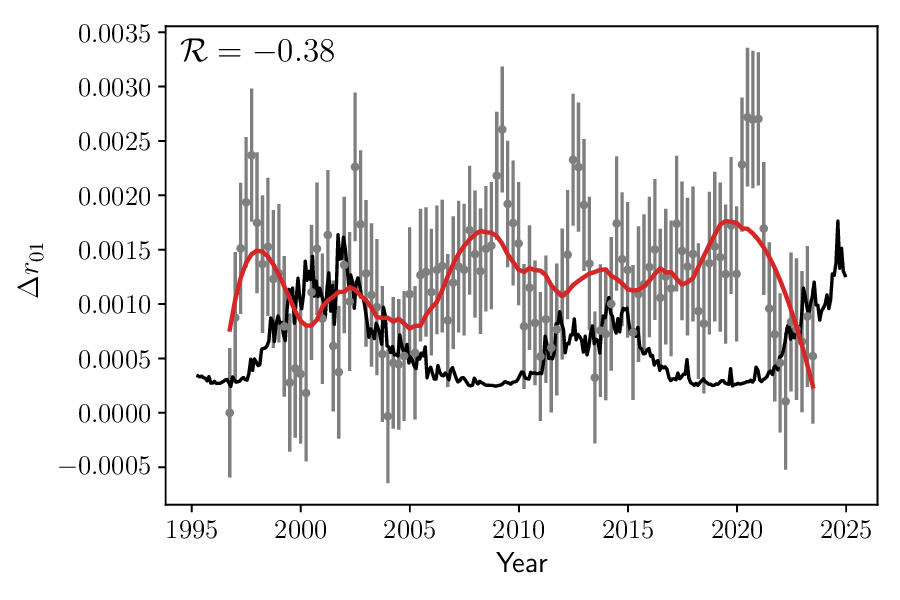} 
  \caption{\centering High-frequency range: 3100 - 3510 $\mu$Hz}
\end{subfigure}
\begin{subfigure}[b]{.49\textwidth}
  \includegraphics[width=.99\linewidth]{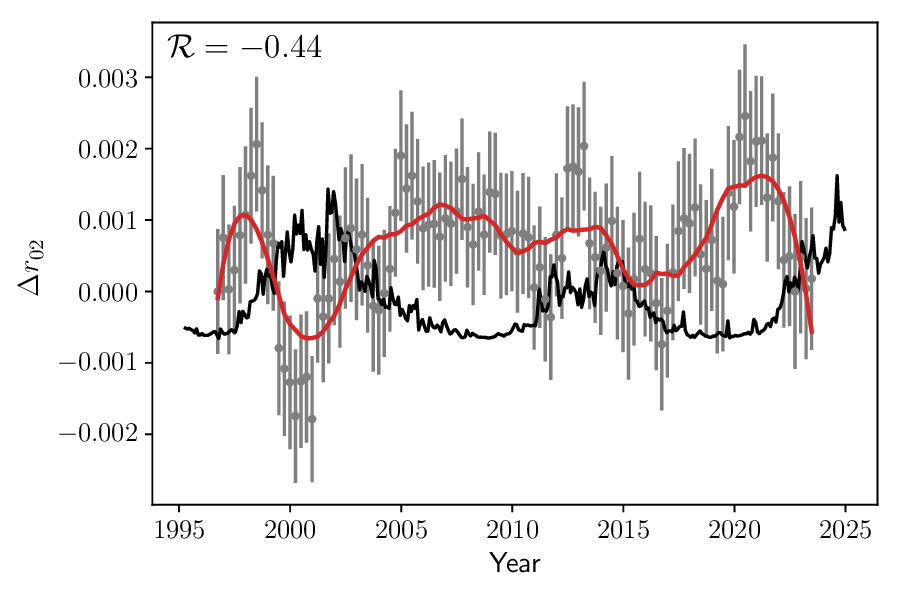} 
   \caption{\centering High-frequency range: 3100 - 3510 $\mu$Hz}
\end{subfigure}
\begin{subfigure}[b]{.49\textwidth}
  \includegraphics[width=.99\linewidth]{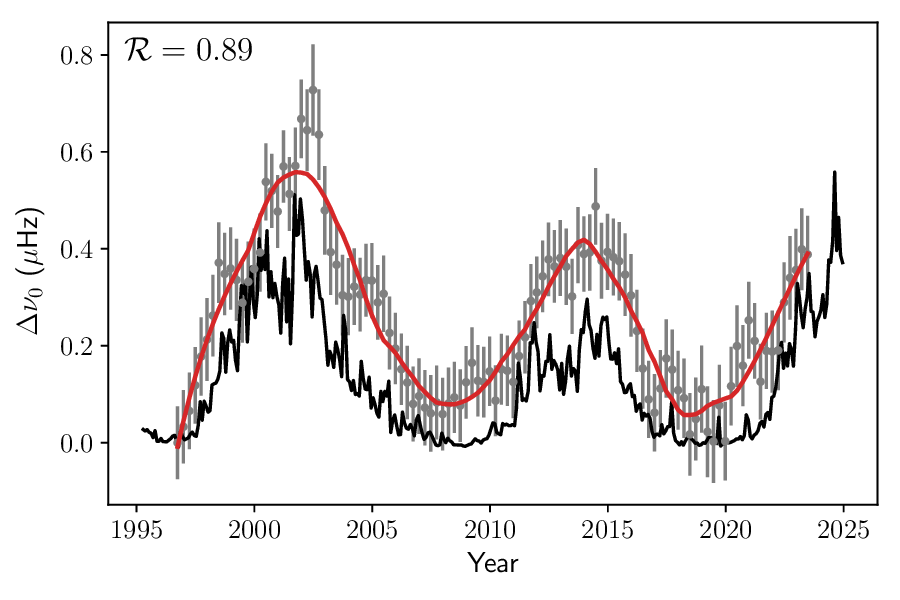} 
   \caption{\centering High-frequency range: 3100 - 3510 $\mu$Hz}
\end{subfigure}}
\caption{Averaged shift due to magnetic activity for frequency separation ratios $r_{01}$ (left column) and $r_{02}$ (middle column), and frequencies of radial modes (right column) observed by the BiSON network. Four frequency ranges were investigated: full range, low-frequency range, medium-frequency range, and high-frequency range (from top to bottom). The red line is obtained by smoothing the grey data with a Savitzky-Golay filter. The 10.7 cm radio emission flux, which serves as a proxy of the solar cycle, is denoted in black. The correlation with the activity proxy is measured with the Spearman rank coefficient $\mathcal{R}$.}
\label{fig:imprint_observations_bison_big_plot}
\end{figure*}

The results for the radial modes align with the extensive literature on this subject \citep[e.g.][for a review]{Broomhall&Nakariakov2015}, indicating that higher frequencies experience greater impacts from magnetic activity. Similarly, as observed with oscillation frequencies, higher radial orders show more pronounced effects from magnetic activity. The shifts in ratios also show a latitudinal dependence, as they depend on the harmonic degree of the frequencies used to construct the ratio. For example, for the $r_{02}$ ratio, the fit of the $l=2$ modes is dominated by the $|m|=2$ peaks, which are more sensitive to low latitudes than the $|m|=0$ component of the nearby $l=0$ modes. Since the magnetic field is also predominantly at lower latitudes, the $l=2$ modes are shifted by more than the $l=0$ modes. This asymmetry in the shift caused by magnetic activity on $l=0$ and $l=2$ modes explains why the $r_{02}$ frequency separation ratio is inefficient at damping magnetic activity, unlike surface effects that have a symmetric nature. Similarly, $l=1$ modes are more shifted than $l=0$ modes, and the $r_{01}$ ratio is also inefficient at damping magnetic activity. For further details, we refer for example to Fig.~B.1 of \citet{Salabert2015}, which illustrates the frequency shift induced by magnetic activity for each harmonic degree during solar cycle 23 and the beginning of cycle 24. Additionally, the asymmetry differs between the $l=0/1$ and $l=0/2$ pairs. Consequently, the $r_{01}$ and $r_{02}$ ratios have different sensitivities to magnetic activity, as observed in our datasets. We note that these observations are consistent with existing literature \citep{Chaplin2005,Thomas2019,Thomas2021}. Figure~\ref{fig:imprint_observations_bison_big_plot} provides a qualitative illustration of the results summarised in Table~\ref{tab:characteristics_observed_shifts} for BiSON data.

\begin{table*}[t]
\centering
\caption{Characteristics of the averaged shifts due to magnetic activity.}
\resizebox{0.99\linewidth}{!}{
\begin{tabular}{lcccccccccccc}
\hline \hline
\multirow{3}{*}{Frequency range} & \multicolumn{6}{c}{BiSON} & \multicolumn{6}{c}{GOLF} \\
\cmidrule[0.4pt](lr){2-7} \cmidrule[0.4pt](l){8-13}
 &  \multicolumn{2}{c}{$\Delta r_{01}$} & \multicolumn{2}{c}{$\Delta r_{02}$} & \multicolumn{2}{c}{$\Delta \nu_{0}$} & \multicolumn{2}{c}{$\Delta r_{01}$} & \multicolumn{2}{c}{$\Delta r_{02}$} & \multicolumn{2}{c}{$\Delta \nu_{0}$} \\
 \cmidrule[0.4pt](lr){2-3} \cmidrule[0.4pt](lr){4-5}  \cmidrule[0.4pt](lr){6-7} \cmidrule[0.4pt](lr){8-9}  \cmidrule[0.4pt](lr){10-11} \cmidrule[0.4pt](l){12-13}
 & $\mathcal{R}$ & $\mathcal{A}$ & $\mathcal{R}$ & $\mathcal{A}$ & $\mathcal{R}$ & $\mathcal{A}$ & $\mathcal{R}$ & $\mathcal{A}$ & $\mathcal{R}$ & $\mathcal{A}$ & $\mathcal{R}$ & $\mathcal{A}$ \\
\hline
Full & $-0.56 \pm 0.11$ & 0.00078 & $-0.63 \pm 0.02$ & 0.00137 & $0.92 \pm 0.03$ & 0.237 & $-0.53 \pm 0.02$ & 0.00068 & $-0.32 \pm 0.05$ & 0.00106 & $0.90 \pm 0.01$ & 0.274 \\ 
Low-frequency & $-0.42 \pm 0.10$ & 0.00087 & $-0.18 \pm 0.08$ & - & $0.19 \pm 0.04$ & - & $-0.25 \pm 0.07$ & - & $0.00 \pm 0.13$ & - & $0.50 \pm 0.04$ & 0.100 \\ 
Medium-frequency & $-0.38 \pm 0.06$ & 0.00112 & $-0.53 \pm 0.09$ & 0.00149 & $0.85 \pm 0.02$ & 0.242 & $-0.49 \pm 0.06$ & 0.00096 & $-0.55 \pm 0.04$ & 0.00125 & $0.89 \pm 0.01$ & 0.211 \\ 
High-frequency & $-0.38 \pm 0.08$ & 0.00152 & $-0.44 \pm 0.05$ & 0.00228 & $0.89 \pm 0.01$ & 0.567 & $-0.37 \pm 0.07$ & 0.00121 & $-0.27 \pm 0.05$ & 0.00183 & $0.86 \pm 0.04$ & 0.541 \\ 
\hline
\end{tabular}}
{\par\small\justify\textbf{Notes.}  The correlation between the averaged shift and the activity proxy, the 10.7 cm radio emission flux, is measured with the Spearman rank coefficient $\mathcal{R}$. The symbol $\mathcal{A}$ denotes the amplitude of the averaged shift, computed by taking the difference between the maximum and minimum of the data smoothed with a Savitzky-Golay filter. The shift amplitude was only evaluated for correlation coefficients strictly larger than 0.25.\par}
\label{tab:characteristics_observed_shifts}
\end{table*}

\section{Supplementary data for the asteroseismic characterisation}
\label{sec:app:supplementary_data_asteroseismic_characterisation}
In Table~\ref{tab:characteristics_MA_on_characterisation}, we present the characteristics of the magnetic activity imprint on the asteroseismic characterisation. The correlation between the stellar parameters and the activity proxy, the 10.7 cm radio emission flux, is measured with the Spearman rank coefficient $\mathcal{R}$. The systematic uncertainty $\sigma_{\mathrm{activity}}$ associated to magnetic activity is evaluated using the methodology described in \citet{Betrisey2025}. A systematic uncertainty is only associated to cases with significant correlation coefficient ($|\mathcal{R}| > 0.25$). The averaged statistical uncertainty $\sigma_{\mathrm{statistical}}$ is computed by averaging the statistical uncertainty of each datapoint.

\begin{table*}
\centering
\caption{Characteristics of the magnetic activity imprint on the asteroseismic characterisation.}
\begin{tabular}{lcccccc}
\hline \hline
 & \multicolumn{3}{c}{BiSON} & \multicolumn{3}{c}{GOLF} \\
 \cmidrule[0.4pt](lr){2-4} \cmidrule[0.4pt](l){5-7}
 & $\mathcal{R}$ & $\sigma_{\mathrm{activity}}$ & $\sigma_{\mathrm{statistical}}$ & $\mathcal{R}$ & $\sigma_{\mathrm{activity}}$ & $\sigma_{\mathrm{statistical}}$ \\
\hline
$n\geq 12$ & & & & & & \\
Mass & $0.39 \pm 0.03$ & 2.6\% & 1.2\% & $0.43 \pm 0.03$ & 2.5\% & 1.2\% \\ 
$Y_0$ & $-0.49 \pm 0.08$ & 10.7\% & 4.6\% & $-0.46 \pm 0.03$ & 10.9\% & 4.3\% \\ 
$\log Z_0$ & $-0.15 \pm 0.19$ & - & 2.7\% & $-0.04 \pm 0.13$ & - & 2.1\% \\ 
Age & $0.62 \pm 0.07$ & 5.1\% & 2.0\% & $0.57 \pm 0.08$ & 4.8\% & 1.8\% \\ 
Radius & $0.39 \pm 0.05$ & 0.9\% & 0.5\% & $0.45 \pm 0.05$ & 0.8\% & 0.4\% \\ 
Mean density & $-0.21 \pm 0.18$ & - & 0.4\% & $-0.22 \pm 0.19$ & - & 0.4\% \\ 
\hline
$n\geq 16$ & & & & & & \\
Mass & $0.30 \pm 0.11$ & 3.3\% & 1.5\% & $0.35 \pm 0.16$ & 3.0\% & 1.5\% \\ 
$Y_0$ & $-0.38 \pm 0.11$ & 11.5\% & 5.4\% & $-0.39 \pm 0.17$ & 11.5\% & 5.3\% \\ 
$\log Z_0$ & $-0.06 \pm 0.15$ & - & 3.0\% & $-0.08 \pm 0.09$ & - & 3.0\% \\ 
Age & $0.61 \pm 0.11$ & 4.3\% & 3.0\% & $0.54 \pm 0.08$ & 4.4\% & 2.9\% \\ 
Radius & $0.30 \pm 0.11$ & 1.2\% & 0.5\% & $0.35 \pm 0.14$ & 1.0\% & 0.5\% \\ 
Mean density & $-0.37 \pm 0.14$ & 0.2\% & 0.4\% & $-0.32 \pm 0.12$ & 0.2\% & 0.4\% \\ 
\hline 
\end{tabular}
{\par\small\justify\textbf{Notes.} The correlation between the stellar parameters and the activity proxy, the 10.7 cm radio emission flux, is measured with the Spearman rank coefficient $\mathcal{R}$. The systematic uncertainty $\sigma_{\mathrm{activity}}$ associated to magnetic activity is evaluated using the methodology described in \citet{Betrisey2025}. The data is considered as uncorrelated if $|\mathcal{R}| \leq 0.25$ and no systematic uncertainty was associated to these cases. The averaged statistical uncertainty $\sigma_{\mathrm{statistical}}$ is computed by averaging the statistical uncertainty of each datapoint. \par}
\label{tab:characteristics_MA_on_characterisation}
\end{table*}

\end{document}